# THE RELATION BETWEEN α-HELICAL CONFORMATION AND AMYLOIDOGENICITY

B. Haimov[1] and S. Srebnik[*]

ABSTRACT

Amyloid fibrils are stable aggregates of misfolded proteins and polypeptides that are insoluble and resistant to protease activity. Abnormal formation of amyloid fibrils *in vivo* may lead to neurodegenerative disorders and other systemic amyloidosis such as Alzheimer's, Parkinson's, and atherosclerosis. Because of their clinical importance amyloids are found under intense scientific research. Amyloidogenic sequences of short polypeptide segments within proteins are responsible for the transformation of correctly folded proteins into parts of larger amyloid fibrils. The α-helical secondary structure is believed to host many amyloidogenic sequences and be a key player in different stages of the amyloidogenesis process. Most of the studies on amyloids focus on the role of amyloidogenic sequences. The focus of this study is the relation between amyloidogenicity and the structure of the amyloidogenic α-helical sequence. We have previously shown that the α-helical conformation may be expressed by two parameters ($\theta$ and $\rho$) that form orthogonal coordinates based on the Ramachandran dihedrals ($\varphi$ and $\psi$) and provide an illuminating interpretation of the α-helical conformation. By performing statistical analysis on α-helical conformations found in the protein data bank, an apparent relation between α-helical conformation, as expressed by $\theta$ and $\rho$, and amyloidogenicity is revealed. Remarkably, random amino acid sequences, whose helical structure was obtained from the most probably dihedral angles as obtained from PDB data, revealed the same dependency of amyloidogenicity, suggesting the importance of α-helical *structure* as opposed to *sequence*.

Alzheimer's, Parkinson's, Creutzfeldt-Jakob's, type II diabetes, atherosclerosis, and prolactinomas, are merely a few examples of diseases caused by the abnormal formation of amyloid fibrils *in vivo* [1]. These fibrils are stable aggregates of misfolded proteins and polypeptides that are insoluble and resistant to protease activity. Many different amyloid fibril structures have been empirically determined [2] with typical diameters ranging from 6 to 12 nm [3], and a cross-β architecture [4] in most cases. The causes for the amyloidogenesis within living organisms vary and include protein synthesis errors, genetic mutations, environmental changes, incorporation of amyloidogenic agents, protein cleavage products, and maturation. Due to their clinical importance, amyloidogenic polypeptides are found under intense scientific research [1,5,6] with a special emphasis on the prediction of amyloidogenic sequences [7–9]. The α-helical secondary structure is believed to be a key player in the process of amyloidogenesis in early self-assembly stages [10], as intermediate structures [11], and because amyloidogenic sequences are often found within helices [12]. In this work, we analyze the structure of amyloidogenic sequences, focusing on the relation between amyloidogenicity and the α-helical conformation.

To understand the dependency of amyloidogenicity on the α-helical conformation, we sampled all of the available α-helical conformations from the rapidly growing world-wide protein data bank (PDB) [13]. An in-house software was developed under C++, TCL, and Matlab$^{TM}$ to perform the processing and the analysis of the PDB data. To build the initial dataset of α-helices, all structures in the PDB were scanned and α-helical conformations were extracted. The determination of α-helical conformations was made according to the predefined structural information included within PDB file headers. Figure 1a presents the distribution of the length of α-helices sampled from PDB in amino acid (AA) units, while Figure 1b presents the distribution of the length of α-helices with unique AA sequences. Comparison shows that only 22.4% of PDB α-helices have a unique AA sequence, while the rest are redundant sequences. Despite the non-negligible percentage of the redundant sequences (77.6%), the distributions of the redundant and the non-redundant sets are very much similar. In both cases we observe that the two most abundant lengths are 6 and 12 AAs, so that the reason for the most abundant α-helical lengths (6 and 12 AAs) is not the redundancy. Studies performed on amyloidogenic sequences tend to focus on the shortest possible peptide sequences that contain all the necessary molecular information for forming typical amyloid fibrils, since longer amyloidogenic sequences may be combined from shorter ones. It was previously shown that peptide sequences as short as pentapeptides [14–16] and even tetrapeptides [16,17] are sufficient for the formation of amyloid fibrils . However, most of the studies performed on amyloidogenic sequences consider hexapeptides as the amount of the known amyloidogenic tetrapeptide and pentapeptide sequences is much fewer than the known amyloidogenic hexapeptide sequences [18–21]. It is not clear from these studies whether the hexapeptide length is of significance or if it is simply the most probable. The distribution of helical lengths presented in Figures 1a and 1b suggests that the significance of amyloidogenic hexapeptides might be the result of their abundance.

The conformation of a polypeptide backbone may be represented using the dihedrals $\varphi$ and $\psi$, plotted on the Ramachandran map [22]. α-Helices are found in the lower left quadrant of the map. Figure 1c presents the distribution of the α-helical ($\varphi,\psi$) conformational pairs that were sampled from PDB. A total of ~15.8e6 conformational pairs were sampled (corresponding to over nearly 2e6 helices), including redundant structures. The reason for leaving redundant phases in this case is because α-helices with similar AA sequences may adopt different conformations [23]. The observed diagonal distribution of ($\varphi,\psi$) pairs is well-known and is the result primarily of steric constraints [24] and hydrogen bonding [25],. However, since ($\varphi,\psi$) pairs do not provide intuitive details regarding the α-helical structure, we utilize an alternative representation using ($\theta,\rho$) pairs that was previously developed [26]. In this orthogonal coordinate system, $\rho$ is the number of residues per turn and $\vartheta$ is the angle of backbone carbonyls relative to the helix direction, thus providing a physical interpretation of the helical conformation. The calculation of ($\theta,\rho$) pairs is done through an approximate linear transformation from the Ramachandran dihedrals. Figure 1d presents the distribution of α-helical ($\theta,\rho$) pairs and clearly demonstrates that the mean α-helical conformation contains approximately 3.6 Res/Turn and a $\theta$ angle of approximately 11°, in good agreement with previous reports [27,28].

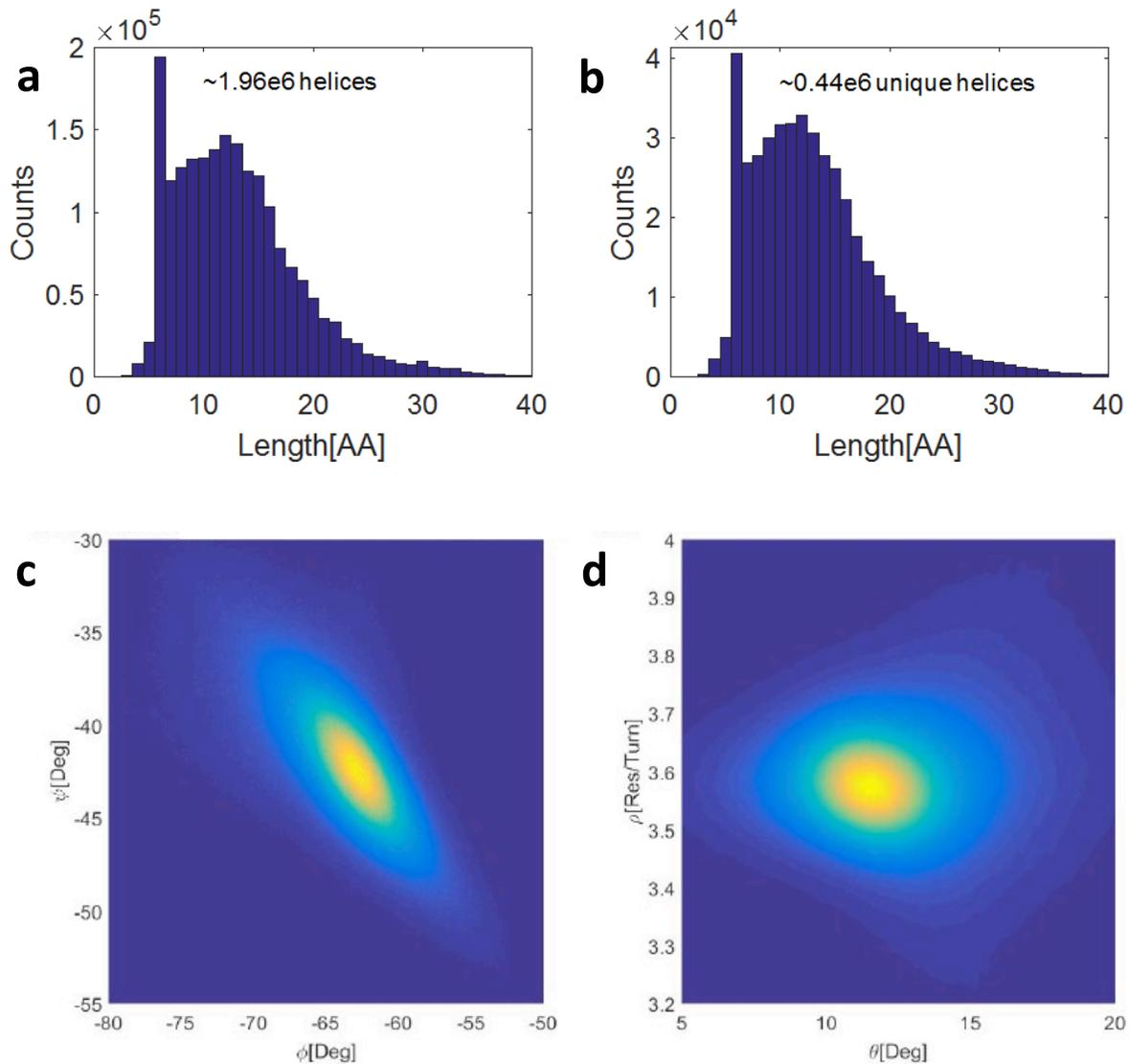

Figure 1: (a) Distribution of the number of AA in all α-helices in the PDB (a total of ~1.96e6). (b) Distribution of α-helices with unique AA sequences (~0.44e6). From the ratio 0.44/1.96 we find that only 22.4% of PDB α-helical sequences are unique, or that 77.6% are redundant. The figures demonstrate that despite the strong redundancy, the distributions remain similar with hexapeptide (6 AA-long) helices as the most abundant. (c) Distribution of PDB α-helical conformation using Ramachandran dihedrals. (d) Distribution of PDB α-helical conformation using ($\theta,\rho$) pairs, where $\theta$ is the angle of backbone carbonyls relative to the helix direction vector and $\rho$ is the number of residues per single α-helix turn. The shades in (c) and (d) represent the abundance of the α-helical conformations, darkest shade for least abundant conformations and brightest shade for most abundant conformations. Darkest blue shade in the background represents zero abundance. A total of ~15.8e6 conformational pairs were sampled from the PDB.

To study the dependency of amyloidogenicity on the α-helical conformation, we prepared three different datasets: (1) α-helices with sequences and conformations taken directly from the PDB, which is the same dataset used for Figure 1; (2) helical hexapeptide sequences taken from the PDB with estimated α-helical conformations (overlapping hexapeptide sequences were considered, i.e., N-5 hexapeptide sequences for a polypeptide with

$N$ amino acid residues); and (3) α-helices with randomly generated hexapeptide sequences and with estimated α-helical conformations. We have previously shown [26] that every transition from one amino acid (AA) to another along the protein backbone has a prominent ($\varphi,\psi$) peak within the α-helix basin which corresponds to a ($\theta,\rho$) peak representing the mean α-helical conformation of the given AA transition, totaling 400 AA-pair transitions (presented in Supplementary Information file Table S1 and Table S2). The ($\theta,\rho$) peak values for each AA pair are used in this study as estimations of the α-helical conformation for the given polypeptide. Thus, by decomposing some given α-helical sequence with $N$ AAs, we get $N$-1 transitions from which we use the peak ($\theta,\rho$) values to estimate the α-helical conformation. For all three datasets, a single representative conformational pair for each polypeptide was calculated as the average ($\theta,\rho$) value of all α-helices with the same AA sequence. Since every hexapeptide α-helix may have variable conformations, it was necessary to estimate a single representative conformation that was calculated as the average conformation of all the α-helices with the same hexapeptide AA sequence. The 1$^{st}$ dataset is used to study the dependency of α-helical conformation on amyloidogenicity for conformations that appear as is in PDB including the redundancy and the difference in α-helical lengths. Averaging over hexamers as opposed to the entire polypeptide chain gave similar results. The purpose of the 2$^{nd}$ dataset is to study the dependency of α-helical conformation on amyloidogenicity for non-redundant hexapeptide sequences, based on mean ($\theta,\rho$) for each AA pair in the sequence. Thus, conformational fluctuations in redundant hexapeptide sequences are not considered. The randomly generated sequences introduced in the 3$^{rd}$ dataset tests for the dependence of amyloidogenicity on sequence (versus structure) and compensates for the potential sequential sparsity of the first two datasets, allowing better stochasticity of the studied sequences.

Previous studies have found that the amyloidogenicity of a polypeptide is a function of its AA sequence [9,29,30]. To find the amyloidogenicity of the helices used in this study we extracted their AA sequences for the three datasets and uploaded it to the online METAMYL web-server [21], which provides an amyloidogenic score based on the AA sequence. For α-helices that contained more than 6 AAs in the 1$^{st}$ dataset, the server was used to calculate the amyloidogenicity sub-score of every hexapeptide sub-sequence and the total amyloidogenicity of the helix was calculated as the average of these sub-scores. By decomposing some given α-helical sequence with $N$ AAs, we get $N$-1 transitions from which we use the prominent ($\theta,\rho$) values to estimate the α-helical conformation. Thus, for every peptide sequence we obtain the set ($\theta,\rho,A$) (A for amyloidogenicity). We binned the data into 10 equally sized bins along the amyloidogenicity axis and calculated the mean ($\theta,\rho,A$) set for each one of the 10 bins. The resulting dependencies of $\theta$ and $\rho$ on the amyloidogenicity for the three datasets are shown as symbols in Figure 2 together with linear regression model for every presented case. Linear regression equations with coefficients of determination of the fitted points are shown in Table 1. Further details and statistical summary of the binned datasets may be found in Supplementary Information Tables S3 to S5. Further information about standard deviations and standard errors of the studied datasets may be found in Supplementary Information file Figures S1 and S2.

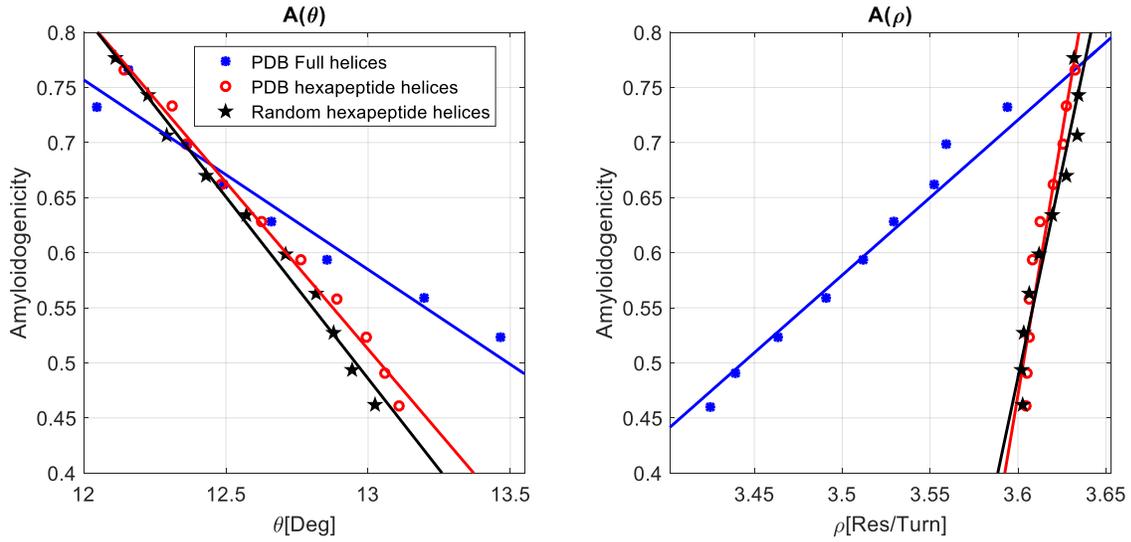

Figure 2: The dependency of amyloidogenicity on $\theta$ and $\rho$ for the three datasets: α-helices with conformations sampled directly from PDB (blue circles), PDB helices broken into hexapeptide α-helices with α-helical conformations estimated from the mean of the strongly peaked ($\theta, \rho$) distribution (red open circles), and randomly generated hexapeptide α-helices with estimated α-helical conformations (black pentacles). The standard error of the mean of every bin is smaller than the marked symbol.

Table 1: Linear regression equations for the dependency of amyloidogenicity on $\theta$ and $\rho$.

|  | A($\theta$) | | A($\rho$) | |
| --- | --- | --- | --- | --- |
|  | Regression line | $R^2$ | Regression line | $R^2$ |
| PDB Full helices | A = 2.82 - 0.17 $\theta$ | 0.97 | A = -4.35 + 1.41 $\rho$ | 0.96 |
| PDB hexapeptide helices[*] | A = 4.45 - 0.3 $\theta$ | 0.99 | A = -33.65 + 9.48 $\rho$ | 0.92 |
| Random hexapeptide helices[*] | A = 4.77 - 0.33 $\theta$ | 0.99 | A = -26.67 + 7.54 $\rho$ | 0.92 |

[*] α-helical conformations were estimated from AA transition data obtained from the PDB [26].

The findings presented in Figure 2 clearly demonstrate a dependency of amyloidogenicity on $\theta$ and evidently a dubious dependency on $\rho$, further discussed below. In all cases, amyloidogenicity decreases with $\theta$ and increases with $\rho$. Conformational fluctuations of redundant hexamer sequences lead naturally to the reduction in the slope of the first dataset relative to the second. In addition, heavily redundant sequences that may appear in dataset 1 will dominate and bias the mean conformation and the overall mean amyloidogenicity score. Since such sequences exist in the 1st dataset (PDB full helices). On the other hand, for the non-redundant datasets (2nd and 3rd) where sequential redundancy was removed, the biasing of the overall mean conformation and the overall mean amyloidogenicity was also removed. Regardless, the trend in each case is self-consistent. The strong similarity between the slopes of the PDB hexapeptides and random hexapeptides is striking and emphasizes the good stochasticity of α-helical hexapeptide sequences found in PDB, but more importantly the role of helical *structure* as opposed to *sequence*.

We have previously shown how $\rho$ and $\theta$ may be influenced by the environment [26]. $\rho$ is sensitive to mechanical forces acting on the α-helix residues and/or the backbone while $\theta$ is sensitive to environmental factors such as pH, ionic strength of surrounding particles, or electric field. Thus, amyloidogenicity may be controlled environmentally as it is proportional to $\rho$ and $\theta$. This conclusion is confirmed by the findings of Fezoui and Teplow [31]. In their study, the authors used trifluoroethanol (TFE) as a co-solvent for amyloid β-protein (Aβ) fibrils. It was clearly shown that by increasing the concentration of TFE, the α-helical phase increased from almost 0% to almost 80%. This way, the authors demonstrated that the propensity for α-helical phases and the propensity for amyloid fibril phases are controllable and dependent on the environment.

Bifurcated hydrogen bonds (BHBs) provide additional steric shielding of the peptide backbone [32] through hydrogen-bonding with water and surrounding molecules that assist in stabilizing the α-helical conformation, and presumably reduces the probability of amyloid formation. Figure 3a presents an illustration of an exposed α-helical backbone with non-BHBs, while Figure 3b presents an illustration of shielded α-helical backbone with BHBs. We therefore assume that amyloidogenicity (*A*) is negatively proportional to the amount of BHBs along the α-helical backbone ($N_{BHB}$):

$$A \propto -N_{BHB} \qquad (1)$$

where the BHB is established between the $i^{th}$ carbonyl oxygen to $i+4^{th}$ amide nitrogen (the main part of the BHB that is required to maintain the α-helical conformation), and between the $i^{th}$ carbonyl oxygen to one of the surrounding molecules (usually water as previously reported by Barlow and Thornton [33]), or to a neighboring side-chain residue.

Careful examination of the findings presented by Barlow and Thornton [33] reveal that when BHBs are established along the α-helical backbone, $\theta$ indeed increases. Therefore, it is plausible to assume that $N_{BHB}$ is also proportional to the angle $\theta$:

$$N_{BHB} \propto \theta \qquad (2)$$

as illustrated in Figure 3b, since $\theta$ (which is the average angle of backbone carbonyls relative to the helix direction normal) must increase when BHBs are established along the α-helical backbone. Combining together relations (1) and (2) results with the following relation:

$$A \propto -\theta \qquad (3)$$

The resulting negative proportionality between the amyloidogenicity *A* and the angle $\theta$, as is observed in Figure 2, suggests that a contribution to the decrease of amyloidogenicity is the establishment of BHBs with the surrounding molecules that apparently lead to the increase of $\theta$ and to the formation of steric shield around the α-helical backbone. Previous efforts have shown that hydrophobic residues and amyloidogenicity are tightly related [12], and that hydrophobic sequences of three or more amino-acids occur less frequently than expected assuming natural selection [34]. Other efforts demonstrated that amyloidogenicity reduces with the increase of the net charge of protein molecules [35–37]. Our findings further suggest that hydrophobic residues which are found near the α-helix backbone will prevent the establishment of BHBs with the α-helical polypeptide backbone that will reduce $\theta$ and increase the amyloidogenicity. Similarly, protein molecules with increased net charge will encourage the establishment of BHBs with the α-helical polypeptide backbone that will eventually increase $\theta$ and decrease the amyloidogenicity.

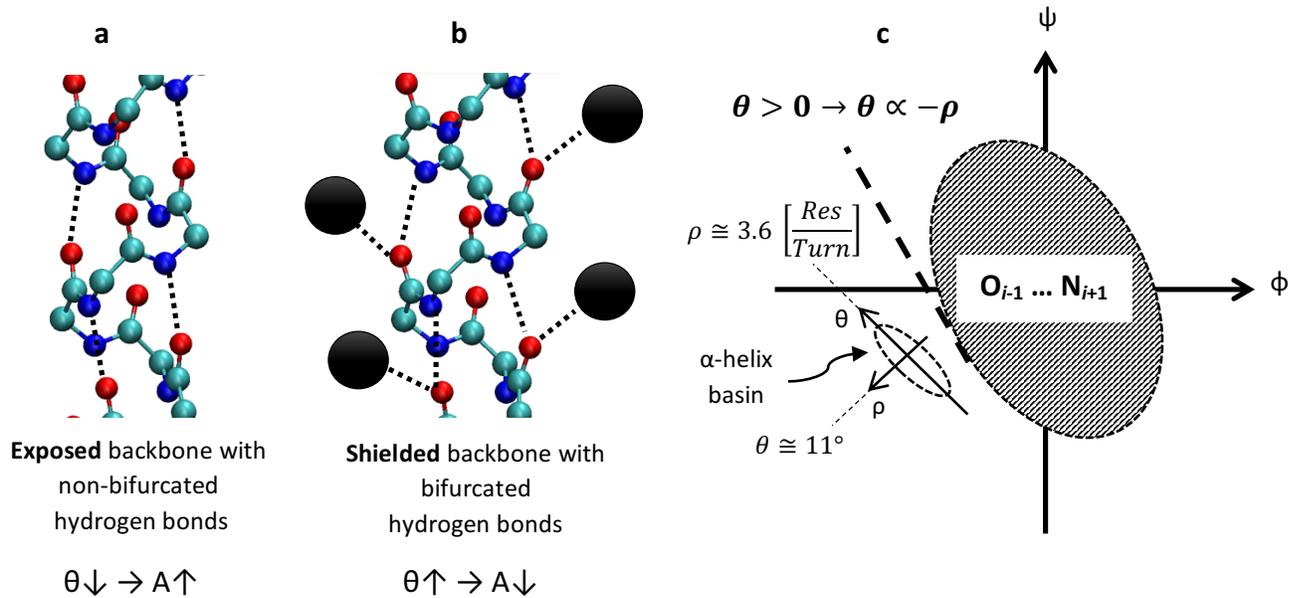

Figure 3: (a) α-helical backbone with non-bifurcated hydrogen bonds and with low $\theta$ results in exposed backbone and higher amyloidogenicity. (b) α-helical backbone with bifurcated hydrogen bonds and with high $\theta$ results in a shielded backbone and lower amyloidogenicity. Dashed lines represent hydrogen bonds. Oxygen, carbon, and nitrogen atoms are shown in red, cyan, and blue, respectively; black circles represent surrounding water molecules or neighboring side-chains that can establish hydrogen bonds with the backbone carbonyls. Residues are not shown for clarity. Illustrations were prepared with VMD [38]. (c) Schematic view of the α-helix basin on the Ramachandran map. The α-helix basin is found near the $O_{i-1}$-$N_{i+1}$ sterically inaccessible region which is due to interactions between $i+1$ amide nitrogen atoms (N) and $i-1$ carbonyl oxygen atoms (O) along the α-helix backbone, where $i$ represents the index of backbone amino acid residues. The tangent line between the sterically inaccessible region and the α-helix basin dictates the negative proportionality between $\rho$ and $\theta$ for $\theta > 0$.

The linear dependency between $A$ and $\rho$ seen in Figure 2 is dubious since $\rho$ is expected to be invariant as there is minimal change in the mechanical forces applied on the helical backbone for the different values of $A$. It is known that the α-helix basin is found near the $O_{i-1}…N_{i+1}$ sterically inaccessible region that is due to steric interference between the $i$-1$^{st}$ carbonyl oxygen and the $i$+1$^{st}$ amide nitrogen atoms (where $i$ represents the index of backbone amino acid residues). As a consequence, some of the α-helical conformations are constrained [25]. To perform qualitative analysis we assume an approximate oval shape of the inaccessible $O_{i-1}…N_{i+1}$ region as depicted in Figure 3c. The tangent line between the α-helix basin and the $O_{i-1}$-$N_{i+1}$ sterically inaccessible region dictates the negative proportion between $\rho$ and $\theta$ for $\theta > 0$. Since $\theta$ is always positive for α-helical conformations [26] and is usually greater than 10°, we deduce that the increase of $\rho$ with increasing amyloidogenicity is due to steric constraints between the α-helix basin and the $O_{i-1}$-$N_{i+1}$ sterically inaccessible region (the slope of the tangent line between the α-helix basin and the $O_{i-1}$-$N_{i+1}$ sterically corresponds with the findings presented in previous reports [24,25]). Hence:

$$\theta > 0 \rightarrow \theta \propto -\rho \rightarrow A \propto \rho \qquad (4)$$

Previous efforts explained that hydrophobicity [12,34] and the net charge of the protein [35–37] are tightly related to amyloidogenicity. Our study provides a complementary explanation of the amyloidogenicity puzzle by relating both hydrophobicity and the net charge to the apparent angle $\theta$ which serves as an indicator for the amount of the established bifurcated hydrogen bonds (BHBs). BHBs are important as they indicate how much the polypeptide backbone is sterically shielded and resistant to conformational changes. An increase in the number of BHBs leads to an increase in $\theta$ and results in a decrease of amyloidogenicity. Moreover, by selecting random AA sequences, we show that the dependence of amyloidogenicity of α-helices on $\theta$ and $\rho$ persist. The fact that these trends are retained for random sequences stresses the role of structure of the α-helix (as is defined by its surrounding environment), as opposed to the sequence of the helix. The interesting relation between amyloidogenicity and the α-helical conformation presented in this study opens novel possibilities for genetic engineering and for the development of innovative drugs.

ACKNOLEDGEMENT

This work was funded in part by the Israel Science Foundation Grant No. 265/16.